\documentclass[aps,prl,twocolumn,groupedaddress,superscriptaddress,amsfonts,amssymb,amsmath,citeautoscript,a4paper]{revtex4-2}

\usepackage{orcidlink}
\usepackage{lipsum}
\usepackage[utf8]{inputenc}
\usepackage[english]{babel}
\usepackage{microtype}
\usepackage{txfonts}
\usepackage{txfontsb}
\usepackage{bbm}
\usepackage{xcolor}
\usepackage{graphicx}
\usepackage{float}
\usepackage{xspace}
\usepackage[skip=0pt, indent=15pt]{parskip}
\usepackage{titlesec}
\titleformat{\paragraph}[runin]{\normalfont\normalsize\bfseries}{}{0}{}[]
\titleformat{\section}[hang]{\normalfont\large\bfseries}{}{0}{}[]
\usepackage{subfiles}
\usepackage{enumerate}
\usepackage[inline]{enumitem}
\usepackage{multirow,rotating}

\usepackage{siunitx}
\DeclareSIUnit[number-unit-product=]\percent{\char`\%} 

\definecolor{lava}{rgb}{0.81, 0.06, 0.13}

\hypersetup{colorlinks,
            linkcolor={lava},
            citecolor={lava},
            urlcolor={lava},
            pdfstartview=FitH}

\makeatletter
\renewcommand{\fnum@figure}{FIG.~\thefigure}
\makeatother

\providecommand{\pnl}[1]{{\textcolor{black}{#1}}}
\usepackage{xifthen}
\usepackage{etoolbox}
\usepackage{soul}

\usepackage{mdframed}
\definecolor{blue-violet}{rgb}{0.54, 0.17, 0.89}

\newmdenv[topline=false, rightline=false, bottomline=false,%
  linewidth=1.25pt, innerrightmargin=0pt, leftmargin=-8pt,%
  innerleftmargin=7pt, skipabove=0pt, skipbelow=0pt,%
  linecolor=blue-violet, fontcolor=blue-violet]{mdleftbar}

\newcommand{\beq}{\begin{equation}\begin{aligned}}
\newcommand{\eeq}{\end{aligned}\end{equation}}

\newcommand{\wse}{WSe\ensuremath{_2}}

\newcommand{\mose}{MoSe\ensuremath{_2}}
\newcommand{\mote}{MoTe\ensuremath{_2}}

\newcommand{\Kplus}{\textsf{\textbf K}}
\newcommand{\Kminus}{\textsf{\textbf K}\ensuremath{'}}

\newcommand{\XD}{\ensuremath{\mathrm{X}_{\mathrm{D}}}}
\newcommand{\Xn}{\ensuremath{\mathrm{X}_{\mathrm{0}}}}
\newcommand{\XT}{\ensuremath{\mathrm{X}_{\mathrm{T}}}}

\newcommand{\polima}{POLIMA -- Center for Polariton-driven Light–Matter Interactions,\\ 
University of Southern Denmark, Campusvej 55, DK-5230 Odense M, Denmark }
\newcommand{\dias}{D-IAS -- Danish Institute for Advanced Study, University of Southern Denmark, Campusvej 55, DK-5230 Odense M, Denmark}
\newcommand{\hbntwo}{International Center for Materials Nanoarchitectonics, National Institute for Materials Science, 1-1 Namiki, Tsukuba 305-0044, Japan}
\newcommand{\hbnone}{Research Center for Functional Materials, National Institute for Materials Science, 1-1 Namiki, Tsukuba 305-0044, Japan}
\newcommand{\auth}{Department of Physics, School of Sciences, Aristotle University of Thessaloniki, 54124 Thessaloniki, Greece}

\begin{document}

\title{Beam Routing through Excitons in Transition Metal Dichalcogenide Monolayers}

\author{Yonas Lebsir\,\orcidlink{0000-0002-9383-0278}}
\thanks{Y.~L. and J.~T.~H. contributed equally to this work.}
\affiliation{\polima}
\author{Jacob Terndrup Heiden\,\orcidlink{0000-0003-4505-6107}}
\thanks{Y.~L. and J.~T.~H. contributed equally to this work.}
\affiliation{\polima}
\author{Jorge Barcia Rodr{\'i}guez\,\orcidlink{0009-0009-1339-804X}}
\affiliation{\polima}
\author{Maria Papadopoulou\,\orcidlink{0009-0005-4772-1731}}
\affiliation{\polima}
\affiliation{\auth}
\author{Kenji~Watanabe\,\orcidlink{0000-0003-3701-8119}}     
\affiliation{\hbnone}
\author{Takashi~Taniguchi\,\orcidlink{0000-0002-1467-3105}}  
\affiliation{\hbntwo}
\author{N.~Asger~Mortensen\,\orcidlink{0000-0001-7936-6264}}
\affiliation{\polima}
\affiliation{\dias}
\author{Sergii~Morozov\,\orcidlink{0000-0002-5415-326X}}
\affiliation{\polima}
\author{Nicolas Ubrig\,\orcidlink{0000-0002-1966-4435}}
\email{ubrig@mci.sdu.dk}
\affiliation{\polima}

\date{\today}

\begin{abstract}
Routing light at the nanoscale typically relies on nanostructured surfaces to imprint directionality on the emission. Using low-temperature, angle-resolved cathodoluminescence spectroscopy, we show that the intrinsic excitonic transitions of a semiconductor can themselves produce routed emission. We probe monolayers of \wse, \mose, and \mote\ and resolve the excitonic species of monolayer \wse\ -- the bright exciton, the trion, and the spin-forbidden dark exciton -- through their distinct angular emission profiles. While the in-plane transition dipoles of the bright exciton and trion radiate predominantly toward the surface normal, the out-of-plane dipole of the dark exciton, inaccessible under normal-incidence optical excitation, produces a directional emission channel at large angles. We further tune the balance between neutral and charged exciton emission through the local dielectric environment. Our results establish dark excitons in TMD monolayers as a platform for directional light emission in compact photonic architectures without additional nanostructuring.
\end{abstract}
\maketitle
Routing of electromagnetic radiation at the nanoscale is a recurring challenge in modern optics, with implications for on-chip photonics, quantum information processing, and sensing technologies. The prevailing approach is to work in the near-field regime, where momentum matching between incoming and outgoing radiation is achieved through gratings, scatterers, antennas, or waveguides~\cite{bharadwaj_optical_2009,gramotnev_plasmonics_2010,yu_flat_2014,coenen_directional_2014,coles_chirality_2016}. A prominent example is the Smith--Purcell effect, where periodically structured surfaces produce directional radiation by a passing electron beam~\cite{smith_visible_1953,yamamoto_interference_2015,dias_active_2026}. Direction-tunable routing has also been realized by combining a structured quantum well with an applied magnetic field, although outcoupling the emission still relies on nanostructured surfaces~\cite{spitzer_routing_2018}. In contrast to these structurally driven platforms, we pursue a fundamentally different route, exploiting intrinsic electronic and optical transitions in semiconductors to achieve beam routing without the need for external nanostructuring.

Monolayer transition metal dichalcogenides (TMDs) are natural candidates for this purpose~\cite{wang_colloquium_2018,mueller_exciton_2018}. These materials exhibit a direct band gap at the inequivalent \Kplus\ and \Kminus\ valleys in the Brillouin zone, and strong spin-orbit coupling further splits the valence band by hundreds of meV and the conduction band by tens of meV, producing a spin-dependent electronic band structure~\cite{xiao_coupled_2012,xu_spin_2014}. This spin dependence is inherited by excitons -- electron-hole pairs bound by Coulomb attraction -- giving rise to a rich variety of bright, spin-allowed, and dark, spin-forbidden, states with distinct transition energies, radiative lifetimes, and polarization selection rules~\cite{ye_probing_2014,wang_colloquium_2018,robert_exciton_2016,selig_excitonic_2016,zhang_magnetic_2017,molas_brightening_2017,dijkstra_ten-valley_2025}. For the spin-forbidden dark excitons (\XD), the transition dipole is oriented out of the basal plane, in contrast to the bright neutral exciton, whose transition dipole lies predominantly within the plane~\cite{schuller_orientation_2013,wang_-plane_2017,scharf_magnetic_2017,molas_probing_2019}. The resulting difference in angular emission profiles makes these excitonic transitions attractive for radiation routing. \wse\ is especially well suited for studying this behavior because \XD\ forms its lowest-energy excitonic state and is therefore spectrally well distinguished~\cite{wang_-plane_2017,feierabend_brightening_2020,robert_measurement_2020}. However, the out-of-plane transition dipole of \XD\ makes it difficult to access through conventional normal-incidence optical excitation, motivating an alternative excitation and detection approach.

Here, we use low-temperature, angle-resolved cathodoluminescence spectroscopy to demonstrate beam routing through the dark exciton state in single atomic layers of TMDs. We investigate several atomically thin TMD monolayers (\mote, \mose, and \wse), and our experiment resolves the relevant excitonic species of the \wse\ monolayer together with their distinct angular emission profiles. Unlike the neutral exciton, \Xn, and the trion, \XT, whose in-plane transition dipoles produce emission strongest around the surface normal, the dark exciton \XD\ exhibits a distinct large-angle emission feature characteristic of an out-of-plane dipole. By engineering the dielectric environment of the supporting substrate, we further demonstrate control over the spectral emission pattern. These results establish dark excitons in TMD monolayers as a viable platform for nanoscale directional light emission and highlight their potential for routing beams into selected channels in compact photonic architectures.
\begin{figure*}[ht]
\centering
\includegraphics[keepaspectratio]{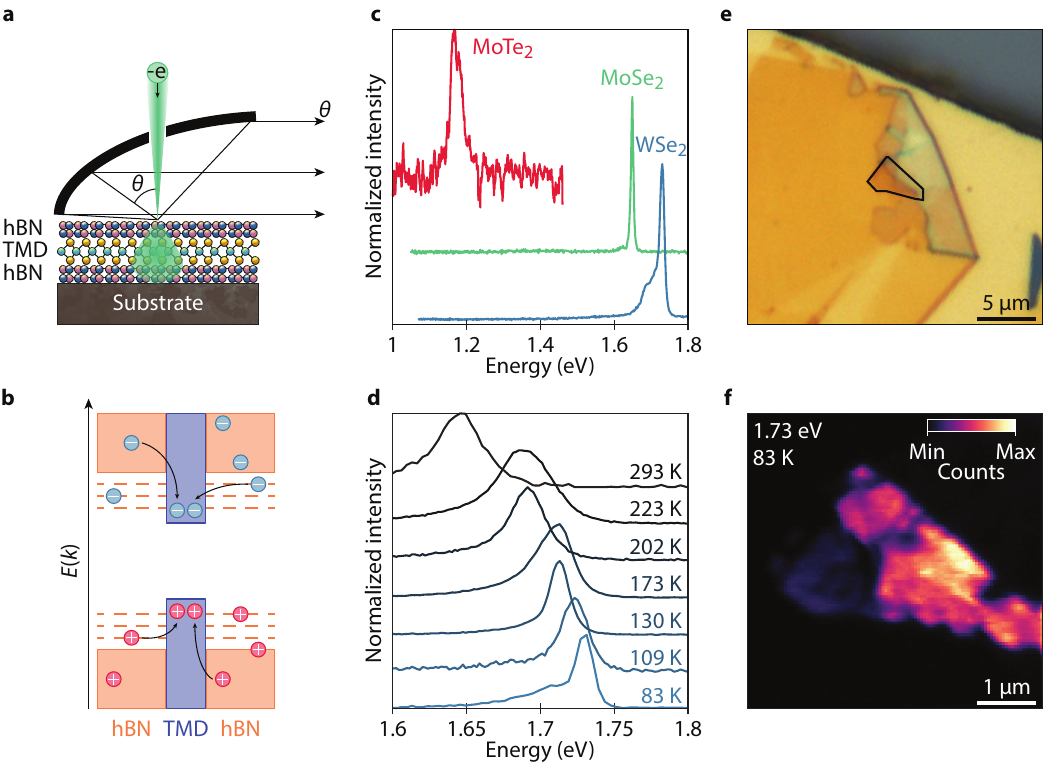}
\caption{
\textbf{Demonstration of electron-beam induced light emission from monolayer TMDs}. \textbf{\pnl{a}.} Schematic of the experimental setup, consisting of a focused electron beam with an accelerating voltage of 5–30~kV and beam currents of 1~nA, focused through a hole in a parabolic mirror with numerical aperture $\mathrm{NA}=0.97$. The collected light is sent to a spectrometer that resolves the emission energy and angle. \textbf{\pnl{b}.} Band diagram illustrating incoherent cathodoluminescence in hBN-encapsulated monolayer TMDs. Electrons and holes are primarily generated in the hBN and relax toward the local band extrema of the TMD, where they form excitons. \textbf{\pnl{c}.} Measured cathodoluminescence spectra at base temperature of hBN encapsulated \mote\ (red line), \mose\ (green line), and \wse\ (blue line) monolayers. \textbf{\pnl{d}.} Angle integrated cathodoluminescence emission spectra of a \wse\ monolayer at different temperatures, showing the expected blue shift and linewidth narrowing upon cooling (curves offset for clarity). \textbf{\pnl{e}.} Optical micrograph of an encapsulated \wse\ monolayer sample placed on a crystalline gold substrate~\cite{boroviks_crystalline_2026, sweedan_2026}. The black solid line delimits the edge of the TMD crystal. \textbf{\pnl{f}.} Cathodoluminescence false-color map of the \wse\ crystal at an emission energy of $1.73$~eV. The reduced integrated intensity is attributed to a locally different thickness of the encapsulating hBN.}
\label{fig1}
\end{figure*}

Figure~\ref{fig1}\pnl{a} schematically illustrates the cathodoluminescence spectroscopy setup. An hBN-encapsulated TMD monolayer is excited by a focused electron beam and the emitted light is collected by a parabolic mirror at base temperatures down to 83~K (see Supporting Information, Sec.~S1). The electron beam can produce both coherent emission, including transition radiation, and incoherent cathodoluminescence~\cite{garcia_de_abajo_optical_2010, polman_electron-beam_2019,ebel_atlas_2025}. The latter is dominant in semiconductors, and arises from electron-hole pairs indirectly generated through inelastic scattering within the pear-shaped interaction volume shown in Fig.~\ref{fig1}\pnl{a}~\cite{zheng_giant_2017,nayak_cathodoluminescence_2019,bonnet_nanoscale_2021,francaviglia_optimizing_2022,taleb_charting_2022,fiedler_photon_2023,bonnet_cathodoluminescence_2024,bonnet_cathodoluminescence_2024}. The encapsulating hBN acts as a wide-bandgap surrounding medium that extends the interaction volume of the electron beam beyond the TMD monolayer. As illustrated by the band diagram in Fig.~\ref{fig1}\pnl{b}, electrons and holes generated in the hBN can diffuse toward the band edge of the TMD layer, where they radiatively recombine~\cite{zheng_giant_2017,nayak_cathodoluminescence_2019,francaviglia_optimizing_2022,ramsden_nanoscale_2023,borghi_cathodoluminescence_2024}. Through its defect network, hBN can even mediate long-range exciton energy transfer in such heterostructures~\cite{darbari_exciton_2026}. Using this approach, we probe the cathodoluminescence response of monolayer excitons over a large energy range, from 1.73~eV for \wse\ and 1.65~eV for \mose\ down to 1.18~eV for \mote\ (Fig.~\ref{fig1}\pnl{c}).%
\begin{figure*}[ht]
\centering
\includegraphics[keepaspectratio]{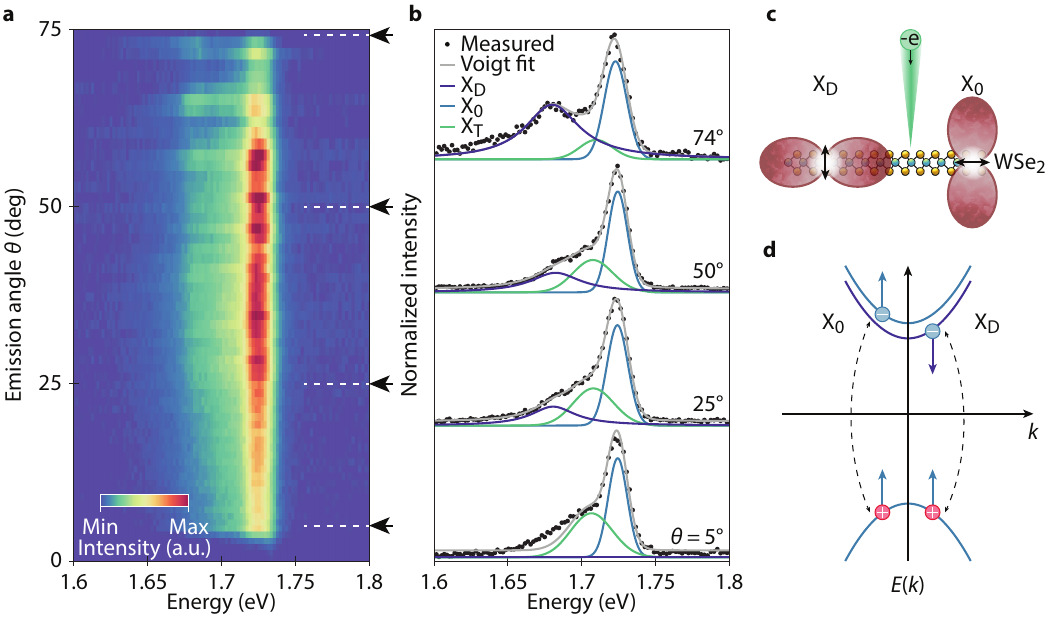}
\caption{
\textbf{Spectral signatures of dark excitons in cathodoluminescence.} \textbf{\pnl{a}.} False-color map of the measured cathodoluminescence intensity, recorded at 83~K on a gold substrate, as a function of emission energy and angle. The dominant emission centered at 1.73~eV originates from the neutral exciton with an in-plane transition dipole. At larger emission angles, a second spectral feature appears, which we attribute to the dark exciton with an out-of-plane transition dipole. \textbf{\pnl{b}.} Cathodoluminescence spectra recorded at the selected emission angles indicated in panel~\textbf{a} (black circles; curves offset for clarity). Solid lines show the fits of the individual excitonic contributions, including the neutral exciton, dark exciton, and trion, shown in blue, purple, and green, respectively. The peak parameters were allowed to vary as fitting parameters within the experimental energy resolution. \textbf{\pnl{c}.} Schematic illustration of the dipole orientations of the dark exciton (\XD) and neutral exciton (\Xn) with respect to the monolayer plane and the incident electron beam. The corresponding dipole emission patterns are sketched, neglecting interference effects from the substrate. \textbf{\pnl{d}.} Scheme of the conduction and valence band extrema of the \wse\ monolayer, around the \Kplus\ point (\Kminus\ is energy degenerate but possesses opposite spins). The bands are spin-split by spin-orbit interaction and give rise to the excitonic structure.
}
\label{fig2}
\end{figure*}

Initial cathodoluminescence measurements reveal strong \Xn\ emission in all three monolayer materials, whereas additional lower-energy emission features are observed exclusively in monolayer \wse. We therefore focus on \wse\ for a detailed investigation, noting that the phenomena reported here are consistently observed across three independently fabricated samples; angle-resolved measurements of two additional samples are shown in Supporting Information, Sec.~S2. Temperature-dependent cathodoluminescence spectra (Fig.~\ref{fig1}\pnl{d}) show that the emission linewidths narrow significantly at low temperature, allowing the lower-energy features to be spectrally resolved. A key advantage of cathodoluminescence is the ability to focus the electron beam to a smaller excitation volume than is possible with diffraction-limited optical probes. This enables spatial variations in emission features, such as the 1.73~eV \Xn\ emission shown in Fig.~\ref{fig1}\pnl{f}, to be resolved with high spatial resolution. Most importantly, the angular resolution of our cathodoluminescence setup (see $\theta$, Fig.~\ref{fig1}\pnl{a})~\cite{osorio_angle-resolved_2016,polman_electron-beam_2019} enables us to resolve the emission directionality of individual excitonic features, and thereby to distinguish exciton species with different transition dipole orientations.

Using this angular resolution, we observe the dark exciton \XD\ in monolayer \wse\ next to the neutral exciton \Xn\ and the trion \XT, as an additional emission feature that appears for emission angles above $\sim 25^\circ$ and becomes increasingly pronounced toward larger angles (Fig.~\ref{fig2}\pnl{a}). We attribute the highest-energy feature at 1.73~eV to \Xn, as it radiates over a broad angular range, consistent with the in-plane transition dipole of the neutral exciton sketched in Fig.~\ref{fig2}\pnl{c}. To systematically separate the overlapping contributions, we fit the spectra with three Voigt profiles while modeling the noise background by polynomials with order $\leq2$, as illustrated in Fig.~\ref{fig2}\pnl{b}. Approximately 20~meV below \Xn, the shoulder at 1.71~eV can be attributed to \XT. Approximately 40~meV below \Xn, the feature at 1.69~eV is attributed to \XD, due to its appearance only at higher emission angles, consistent with the \XD\ out-of-plane transition dipole (Fig.~\ref{fig2}\pnl{c}). The lower energy of \XD\ relative to \Xn\ is consistent with the band structure of monolayer \wse\ (Fig.~\ref{fig2}\pnl{d}), and the measured splitting of approximately 40~meV agrees with established values~\cite{wang_-plane_2017,zhang_magnetic_2017,wang_colloquium_2018} for the neutral-dark exciton splitting (in contrast to monolayer MoX$_2$, X = S, Se, Te, where \XD\ lies above \Xn). Under surface-normal optical excitation, only the in-plane dipole of the neutral exciton is optically accessible, whereas the out-of-plane dipole associated with the dark exciton is forbidden. Electron-beam excitation, by contrast, can couple to both dipole orientations, as it relies on carrier transfer from hBN and not optical selection rules. The measurements therefore reveal a directional cathodoluminescence channel arising from the out-of-plane dark exciton dipole in monolayer \wse.

\begin{figure*}[ht]
\centering
\includegraphics[keepaspectratio]{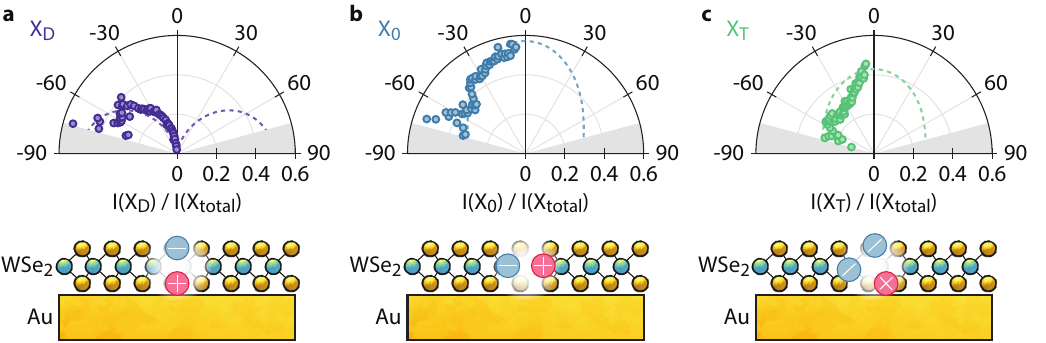}
\caption{
\textbf{Angular emission profile of distinct exciton species.}
\textbf{\pnl{a}.} Polar plot of the dark exciton emission profile (\XD, purple circles). The radial coordinate represents the normalized intensity, I(\XD)/I($\mathrm{X}_\mathrm{total}$), with the normalization used to account for the instrument response, as described in the main text. The dashed line shows a simplified simulation of the emission pattern of the structure shown below, where the hBN layers are omitted for clarity but included in the calculation, confirming the predominantly out-of-plane character of the emission. The gray shaded area indicates the cutoff angle of the parabolic mirror. \textbf{\pnl{b}.} Same as in \textbf{\pnl{a}}, but for the neutral exciton (\Xn, blue circles) with an in-plane dipole moment, as sketched below. The dashed line shows that the dominant emission is directed normal to the basal plane of the crystal. \textbf{\pnl{c}.} Polar emission profile of the trion (\XT, green circles), following the same analysis as in \textbf{\pnl{a}} and \textbf{\pnl{b}}. Its dominant dipole transition is inherited from the neutral exciton, explaining the emission pattern concentrated near the surface normal of the \wse\ plane.
}
\label{fig3}
\end{figure*}

\begin{figure}[ht!]
\centering
\includegraphics[keepaspectratio]{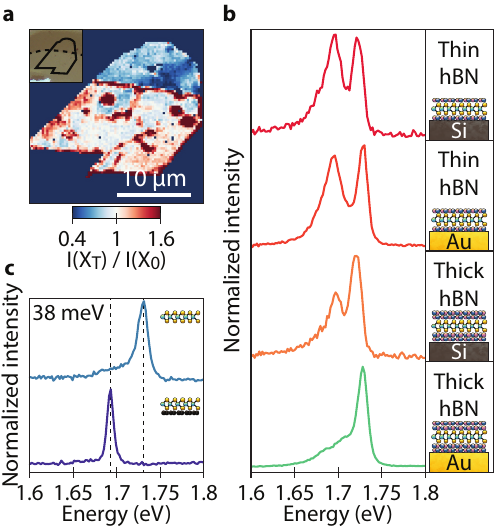}
\caption{
\textbf{hBN thickness dependency.}
\textbf{\pnl{a}.} Cathodoluminescence false-color map, recorded at 83~K, of the trion-exciton intensity ratio in a sample where the same \wse\ is placed on top of two distinct hBN thicknesses ($\approx$ 20 and 50~nm). The inset displays an optical micrograph of the sample with a black outline of the monolayer. The intensity ratio clearly correlates with the hBN thicknesses. \textbf{\pnl{b}.}~Cathodoluminescence spectra of different samples (sketched on the right next to each respective spectrum) with varying hBN thicknesses and substrates. Every spectrum is normalized with respect to its own maximum.
\textbf{\pnl{c}.} Maximum normalized cathodoluminescence spectra of a \wse\ monolayer encapsulated in hBN, compared to the same \wse\ crystal in contact with graphene.
}
\label{fig4}
\end{figure}

It is remarkable that the dark exciton emission can be resolved at temperatures as high as 100~K (see Supporting Information, Sec.~S3). Because \XD\ has a comparatively long lifetime ($\sim 10^2$~ps), it is susceptible to phonon-assisted scattering into the much shorter-lived neutral state ($\sim 1$~ps), especially given phonon modes around 40~meV in \wse~\cite{robert_fine_2017}. As a result, the dark exciton peak is progressively suppressed at higher temperature. Lower temperatures would suppress this relaxation channel and further sharpen the spectral features. Nevertheless, spectral fitting still allows all excitonic contributions to be reliably identified under the present conditions.

The distinct transition dipole orientations of the three excitonic species is expected to produce characteristic and potentially opposing emission profiles, which we quantify across the measured angular range. To compare their angular dependence, we determine the fitted intensity $I(j)$ of each resonance and calculate its normalized spectral weight $w(j)$ at every collection angle,
\[
w(j)=\frac{I(j)}{I(\Xn)+I(\XT)+I(\XD)},
\]
where $j=\Xn,~\XT,~\XD$. This normalization removes angle-dependent variations in the overall detection efficiency that affect all three resonances equally. The resulting polar plots are shown in Fig.~\ref{fig3}. The relative contribution of \Xn\ (Fig.~\ref{fig3}\pnl{b}) is largest near the surface normal and decreases toward larger collection angles. This is expected, as the toroidal emission of the in-plane dipole is zero at $\theta=90^\circ$ for all dipoles parallel to the collection path. \XT\ (Fig.~\ref{fig3}\pnl{c}) follows a similar trend to \Xn, with a slightly more isotropic distribution, potentially reflecting recoil effects associated with the additional charge. In contrast, the dark exciton \XD\ (Fig.~\ref{fig3}\pnl{a}) exhibits the opposite behavior: its relative intensity increases strongly toward large emission angles. These distinct angular profiles are consistent with the dipole-emission patterns illustrated in Fig.~\ref{fig2}\pnl{c} and computed using a simplified simulation of the stack under investigation (dashed lines in Fig.~\ref{fig3})~\cite{schneider_direct_2020,akerboom_angle-resolved_2025}. Because the hBN thickness is much smaller than the emission wavelength, interference effects remain modest in the present geometry. Taken together, these results show that cathodoluminescence can distinguish excitonic species in TMD monolayers not only by their spectral signatures, but also by their characteristic angular emission profiles, establishing transition dipole orientation as a mechanism for exciton-selective nanoscale beam routing. To our knowledge, this constitutes the first identification of individual excitonic species from the angular distribution of their cathodoluminescence which opens a route to controlling beam routing at the nanoscale through dipole orientation and emission interference.

We next show that the emission routes can be tuned by the interaction of the electron beam with the local dielectric environment. The cathodoluminescence intensity of each excitonic species depends on how efficiently the electron beam generates free carriers in the surrounding hBN, which subsequently relax into the \wse\ monolayer, as introduced in Fig.~\ref{fig1}\pnl{b}. Such effect is already apparent in the cathodoluminescence map of Fig.~\ref{fig1}\pnl{f}, where the emission intensity varies across regions with different hBN thicknesses. To isolate this dependence, we examine another hBN-encapsulated monolayer \wse\ sample in which the bottom hBN flake contains two regions of visibly different thickness (see inset in Fig.~\ref{fig4}\pnl{a}), while the top encapsulation remains unchanged. Fig.~\ref{fig4}\pnl{a} shows the corresponding ratio between the trion and neutral exciton emission, $I(\XT)/I(\Xn)$, which reveals a clear spatial contrast that correlates with the hBN thickness variation, while the optical room-temperature photoluminescence is virtually homogeneous (see Supporting Information, Sec.~S4). This suggests that the encapsulating layer thickness influences the efficiency with which different excitonic species are generated and recombine. We further compare \wse\ cathodoluminescence spectra from different sample configurations in Fig.~\ref{fig4}\pnl{b}, including thin and thick hBN encapsulation on silicon and gold substrates~\cite{boroviks_crystalline_2026,sweedan_2026}. Across these configurations, the trion contribution decreases systematically from thin-hBN samples toward thicker hBN. Monte Carlo simulations support this trend, showing more electron scattering, and thus potentially more efficient carrier generation, for thinner hBN encapsulation (see Supporting Information, Sec.~S5). These results establish the local dielectric environment, and hBN thickness in particular, as a passive control parameter for the balance between neutral and charged excitonic cathodoluminescence.

Whereas the hBN thickness influences the generation and transfer of beam-induced carriers, an adjacent graphene layer provides a complementary means of controlling their availability within the \wse\ monolayer. The trion peak is populated when excess free carriers are available to bind to the exciton, so the trion-to-exciton ratio directly reflects the local carrier density. Electrostatic gating is the conventional way to steer this balance, injecting or depleting carriers with an applied bias~\cite{wu_probing_2026}. A similar effect can be obtained in graphene-contacted regions (Fig.~\ref{fig4}\pnl{c}) where the trion emission is strongly suppressed, showing how graphene acts as a passive gate that brings the monolayer near charge neutrality~\cite{lorchat_filtering_2020}. At the same time, the \Xn\ emission in the graphene-contacted regions is redshifted by 38~meV relative to the hBN-only regions. This shift is consistent with the modified dielectric screening introduced by graphene, and may also be influenced by charge transfer or nonradiative coupling to graphene~\cite{raja_coulomb_2017}. These measurements show that the local environment provides an effective handle for reshaping the cathodoluminescence spectrum of monolayer \wse. In particular, hBN thickness, substrate material, and nearby graphene layers can tune the balance between neutral and charged excitonic emission.

In summary, we have used an electron beam to excite the different exciton species in the TMD monolayers \wse, \mose, and \mote\ and determined their species-resolved angular emission profiles. In contrast to conventional routing strategies that rely on nanostructuring, our results demonstrate that these intrinsic electronic transitions in semiconductors provide a viable platform for routing light emission. Substrate engineering provides a handle to adjust both the emission angle and the emission energy of the quasi-particle. This control may be exploited further by coupling to polaritons and harnessing near-field effects, for instance in waveguide geometries. Finally, our work demonstrates that cathodoluminescence successfully excites different quasi-particles in solids, an approach that can be extended, with its high spatial resolution, to single-photon emitters or individual moiré sites in 2D materials and van~der~Waals heterostructures.

\noindent\textbf{Supporting Information:} Additional experimental details on sample fabrication and cathodoluminescence spectroscopy, angle-resolved cathodoluminescence of additional samples, temperature dependence of the dark exciton, photoluminescence characterization, and Monte Carlo simulations of electron scattering in hBN-encapsulated samples (PDF).

\section{Acknowledgments}

The Center for Polariton-driven Light-Matter Interactions (POLIMA) is funded by the Danish National Research Foundation (Grant No.~DNRF165). N.~U. is supported by the Novo Nordisk Foundation (Grant No. NNF25OC0103142). The work presented here is supported by the Carlsberg Foundation (Grant No. CF24-2081) and the VILLUM Foundation (Grant No. 16498).

\section{Author contributions statement}

Y.~L. and J.~T.~H. contributed equally to this work.
Y.~L., J.~T.~H., J.~B.~R., and M.~P. assembled and structurally characterized the samples.
Y.~L. and J.~T.~H. performed the measurements.
N.~U., S.~M., and N.~A.~M. supervised the research.
The hBN crystals were grown by K.~W. and T.~T.
All authors participated in the analysis of the data and the writing of the manuscript.

\section{Competing interests statement}

The authors declare no competing interests.

%

\clearpage
\end{document}